\documentclass[12pt]{article}
\usepackage{amsmath}
\usepackage{amsthm}
\usepackage{bbm}
\usepackage{float}
\usepackage{graphicx}
\usepackage{enumerate}
\usepackage{natbib}
\usepackage{subfig}
\usepackage{tikz}
\usepackage{url} 

\theoremstyle{plain}
\newtheorem{assumption}{\sc Assumption}

\newtheorem{theorem}{\bf Theorem}

\begin{document}

\def\spacingset#1{\renewcommand{\baselinestretch}%
{#1}\small\normalsize} \spacingset{1}
\def\t{{ \mathrm{\scriptscriptstyle T}}}
\def\v{{\varepsilon}}
\def\pr{p}
\def\f{f}
\def\expit{\operatorname{expit}}
\def\logit{\operatorname{logit}}
\def\OR{{\rm  OR}}
\newcommand\independent{\protect\mathpalette{\protect\independenT}{\perp}}
\def\independenT#1#2{\mathrel{\rlap{$#1#2$}\mkern2mu{#1#2}}}
\makeatletter
\newcommand*{\ind}{%
	\mathbin{%
		\mathpalette{\@ind}{}%
	}%
}
\newcommand*{\nind}{%
	\mathbin{
		\mathpalette{\@ind}{\not}
	}%
}
\newcommand*{\@ind}[2]{%
	\sbox0{$#1\perp\m@th$}
	\sbox2{$#1=$}
	\sbox4{$#1\vcenter{}$}
	\rlap{\copy0}
	\dimen@=\dimexpr\ht2-\ht4-.2pt\relax
	\kern\dimen@
	{#2}%
	\kern\dimen@
	\copy0 
} 
\makeatother


  \title{\bf A self-censoring model for multivariate nonignorable nonmonotone missing data}
  \date{}
    \author{Yilin Li, Wang Miao\\
   Department of Probability and Statistics, Peking University\\
    Ilya Shpitser \\
    Department of Computer Science, Johns Hopkins University \\
    and\\
    Eric J. Tchetgen Tchetgen\\
    Department of Statistics and Data Science, \\The Wharton School of the University of Pennsylvania}

\maketitle

\begin{abstract}
We introduce a self-censoring model    for multivariate  nonignorable nonmonotone  missing data, where the missingness process of each outcome is affected by its own value and is associated with missingness indicators of other outcomes,  while conditionally independent of   the other outcomes.
The self-censoring model complements  previous graphical approaches for the analysis of multivariate    nonignorable missing data.
It is identified under a completeness condition stating that any variability in one outcome can be captured by variability in the other outcomes among complete cases.
For estimation, we propose   a suite of semiparametric estimators including doubly robust estimators that deliver valid inferences under partial misspecification of the full-data distribution. 
We evaluate the performance of the proposed estimators with   simulations  and apply them to analyze a study about the effect of highly active antiretroviral therapy on preterm delivery of HIV-positive mothers.
\end{abstract}

\noindent%
{\it Keywords:}  Doubly robust estimation; Identification; Missing not at random; Multivariate missingness;  Nonmonotone missingness.

\spacingset{1.75}
\section{Introduction}

Missing data   arise  ubiquitously in empirical studies.
The  missing data  mechanism  is called  missingness at random or ignorable if it does not depend on the missing variables conditional on fully-observed ones. 
Otherwise, it is called missingness  not at random or nonignorable. 
Nonignorable missingness introduces fundamental  challenges to identification, where identification means that the full-data distribution or a statistical functional of interest is uniquely determined from the observed-data distribution. 
For a single missing outcome, \cite{miao2016identifiability,wang2014instrumental} illustrated  that identification is not ensured even for fully parametric models, 
and fully-observed auxiliary variables such as  shadow variables \citep{miao2016on,d2010new} or  instrumental variables \citep{liu2020identification,sun2018semiparametric} have been used to  achieve identification under nonignorable missingness. 
However, for multivariate missing data, the missingness   of each variable may depend on different partially-observed variables and the missingness of other variables, which leads to   difficulty for  specifying provably identifiable models.
For multivariate missing data,  the missingness is called  monotone if the variables can be ordered such that  once a variable is unobserved, all variables later in the order are also missing. This occurs frequently in longitudinal studies  where missingness is due to dropout.
Otherwise,  arbitrary patterns of missingness can arise, which is called nonmonotone missingness.

Previous researchers have studied a variety of  models for  nonignorable nonmonotone missingness,    including the group permutation model \citep{robins1997non} which states that missingness can depend on ``previous'' and ``observed future''   but not on the present outcome, 
and the block-sequential missing at random model \citep{zhou2010block} which characterizes the potentially nonignorable missingness for blocks  of variables  in a sequential way. 
\cite{tchetgen2018Discrete} introduced discrete choice models which connect the multinomial missingness mechanisms to some underlying utility functions.
\cite{linero2017bayesian} proposed a nearest identified pattern restriction for identification under Bayesian  inference framework.  
Directed acyclic graphs are a useful tool for describing   multivariate missingness   and studying identification.
\cite{fay1986causal,ma2003identification} studied identification  of directed acyclic graphs with categorical  outcomes. 
\cite{mohan2021graphical} provided a comprehensive  review on  missing data methods using graphical models,
established   conditions for  identification   and proposed   testable implications for directed acyclic graphs. 
\cite{nabi2020full} established necessary and sufficient   identifying conditions  for the full-data distribution when the only restrictions on the full-data distribution are implied by its factorization with respect to a directed acyclic graph.
Notably, the  no self-censoring model was recently proposed for the  analysis of multivariate nonignorable nonmonotone missing data, 
which assumes that the missingness process of each outcome does not depend on its value after conditioning on the other outcomes and missingness indicators.
\cite{sadinle2017itemwise,shpitser2016consistent,malinsky2021semiparametric} showed that  nonparametric identification can be  achieved under the no self-censoring model.
\cite{malinsky2021semiparametric} developed semiparametric efficiency theory and doubly robust estimation under the no self-censoring model.

In contrast to the no self-censoring model,  it   is commonly encountered   in practice that
the missingness of an outcome is affected by its own value.
This  is plausible  in situations where  the failure of data collection,  data loss, or units' unwillingness to respond is affected  by data values.
Examples include religious beliefs and sexual preferences in epidemiological studies,
smokers not reporting their  smoking behavior in insurance applications, 
and voters not disclosing their political preferences in  election surveys.
Although this notion of self-censoring  for a univariate outcome  has been studied extensively \citep[e.g.,][]{d2010new,wang2014instrumental,miao2016identifiability,sun2018semiparametric},  the multivariate  self-censoring mechanism   is  rarely studied with only few exceptions.
\cite{brown1990protecting} considered a self-censoring mechanism for multivariate normal outcomes with a  fully-observed outcome.

In this paper, we describe a  self-censoring model for multivariate  nonignorable nonmonotone missing data.
The self-censoring model reveals that  the missingness process of each outcome can depend on its underlying value,  the missingness status of other outcomes, and some possibly fully-observed covariates, but not on other partially-observed outcomes. 
We establish the identification condition for the full-data distribution and develop a suite of semiparametric estimation methods including a doubly robust one.
We do not impose any parametric restriction but invoke  a completeness condition that is commonly used in nonparametric identification problems, 
which accommodates both discrete and continuous variables. 
A generalized odds ratio parameterization is adopted to  facilitate estimation by factorizing the full-data distribution into a baseline propensity score model, a baseline outcome model, and an odds ratio function.
Under this parameterization,  we   develop an inverse probability weighted estimator and an outcome regression/imputation-based estimator, 
which require correct specification of the respective  baseline propensity score or the baseline outcome model, together  with the odds ratio function. 
To promote robustness against model misspecification, we propose a doubly robust estimator, which is consistent as long as the   odds ratio model is correct and at least one of the baseline  models is correct but not necessarily both.
We evaluate performance of the proposed estimation methods with simulations
and apply them to a real data problem about the effect of highly active antiretroviral therapy on preterm delivery of mothers with HIV-positive and maternal hypertension in Botswana. 
Our  analysis results show that the data  are   likely to be prone to nonignorable missingness, and highly active antiretroviral therapy increases the risk of preterm delivery.
All proofs are relegated to   the Supplementary Material.

\section{The Self-censoring model}\label{sec:2}
\subsection{Model assumptions}\label{subsec:21}  
Let $Y=(Y_1,\ldots, Y_p)$ denote a vector of variables subject to missingness, $R=(R_1,\ldots, R_p)$     the vector of missingness indicators  or patterns with $R_{i}=1$ if $Y_{i}$ is observed and $R_{i}=0$ otherwise,  and $X$   a  vector of $d$ possibly fully-observed covariates. 
We use capital letters for random variables and lowercase letters for realization of the  corresponding variables.
The observed data are comprised of independent and identically distributed realizations of $(Y_{(R)}, R, X)$ where $Y_{(R)} = (Y_{i}: R_{i}=1)$ denotes the observed variables of $Y$ for pattern $R$. 
Let $\mathcal{R}=\{r \in \{0,1\}^{p}:\pr(r)>0\}$ be the set of all  missingness patterns.  
We use  $R = r$ as shorthand for $\left(R_{1}, \ldots, R_{p}\right)=\left(r_{1}, \ldots, r_{p}\right)$,  $R=1$ for $R=(1,\dots,1)$, and likewise $R=0$ for $R=(0,\dots,0)$. 
For  a  vector $V$, let $V_{-i}$ denote the subvector  after removing $V_i$; let $R_{<i} = (R_{1},\dots,R_{i-1})$ and likewise  $R_{>i}=(R_{i+1},\dots,R_{p})$.
We let $\f(\cdot)$ denote a probability density function induced by some dominating measure $\nu$.
Given two  missingness patterns $r, r^{\prime} \in \{0,1\}^p$, we denote $r \preceq r^{\prime}$ if $r_i \leq  r_{i}^{\prime}$ for all $i$.
This defines a partial order  of the missingness patterns that  all observed outcomes in pattern $r$ are also observed in pattern $r'$. 
The following assumption characterizes a self-censoring mechanism for multivariate missing data. 
\begin{assumption}\label{assump:self}
\begin{minipage}[t]{13cm}
\begin{enumerate}
\item[(i)] Self-censoring: For $i =  1,\dots, p$, $R_{i}\ind Y_{-i} \mid (X,Y_{i}, R_{-i})$;
\item[(ii)]  Positivity: For any $r\in \mathcal{R}$, $\pr(r \mid x,y) > c$ almost surely for some constant $c>0$, and if $r \preceq r^{\prime}$ then $r^{\prime} \in \mathcal{R}$.
\end{enumerate}
\end{minipage}
\end{assumption}
Assumption \ref{assump:self} describes  a  nonignorable missingness mechanism  for multivariate outcomes.
Condition (i) reveals a self-censoring mechanism  that the  missingness   of each outcome is affected by its value,  
missingness indicators of other outcomes, and fully-observed covariates,  but not affected by  the value of  other outcomes. 
A special case is that  $R_{i}\ind (Y_{-i},R_{-i}) \mid (X,Y_{i})\ $ for $i =  1,\dots, p$,  which characterizes a \textit{strict self-censoring model} where the missingness of each outcome can only depend on the outcome's value and fully-observed covariates, also known as the generalized censoring mechanism by \cite{brown1990protecting} or self-masking by \cite{mohan2018linear}. 
The self-censoring mechanism is plausible particularly in surveys with  certain sensitive questions (e.g., HIV status, income, or drug use)  so that social stigma makes the nonresponse of outcomes directly dependent on their underlying values. 
Condition (i) imposes restrictions on the observed-data distribution. Therefore,  some of the conditional independence can in principle be refuted with the observed data \citep{d2010new}. We let $R_{-i}^{o} = \{R_j: R_j=1, j\neq i\}$, $R_{-i}^{m} = \{R_j: R_j=0, j\neq i\}$, and $Y_{-i}^{o} = \{Y_j: R_j=1, j\neq i\}$ for the observed outcomes other than the $i$th. Consider the following equation in $\phi(\cdot)$ which only involves the observed data,
\[
E\left\{\frac{R_i}{\phi(X, Y_i)}\mid X, Y_{-i}^o, R_{-i}^o=1,  R_{-i}^m=0\right\} = 1.
\]
Under condition (ii), $R_{i}\ind Y_{-i}^o\mid X, Y_{i}, R_{-i}^o=1, R_{-i}^m=0$ can be rejected if  there is no solution of $\phi(\cdot)$ belongs to $(0,1]$ for $i=1,\dots,p$,  while others such as $R_{i}\ind Y_{-i}^m\mid X, Y_{i}, R_{-i}^o=1, R_{-i}^m=0$ are untestable with the observed data distribution.

 The first part of condition (ii) is a common positivity condition stating that  the propensity for any missingness pattern is  bounded away from zero. It is necessary for identification of  the full-data distribution and consistent estimation of its functionals. If a missingness pattern exists, the second part of  (ii) requires the existence of  missingness patterns  with fewer unobserved outcomes; this condition rules out certain missingness mechanisms such as monotone missingness. 
To illustrate the necessity of (ii), suppose two binary outcomes $Y_{1}$ and  $Y_{2}$  are missing concurrently, i.e., $R_{1} = R_{2}$, then  condition (i) holds but (ii) fails because the patterns $(r_1=1,r_2=0)$ and $(r_1=0,r_2=1)$ do not exist. In this case, the full-data distribution
is not identifiable because  $\f(y_1,y_2,r_1=r_2=0)$ cannot be recovered from the observed data.

A notable model for multivariate nonignorable missing outcomes is  the no self-censoring model previously studied by \cite{sadinle2017itemwise,shpitser2016consistent,malinsky2021semiparametric}, which assumes that $R_{i}\ind Y_{i} \mid (X,Y_{-i},R_{-i})$ for $i =  1,\dots, p$, 
i.e.,  the missingness of each outcome is not influenced by  its own value but by the remaining partially-observed outcomes.
The model in Assumption \ref{assump:self} complements by considering the self-censoring mechanism. 
It has been noted that  if self-censoring  exists the full-data distribution is unlikely to be  identified  without additional restrictions  \citep[e.g.,][]{nabi2020full,mohan2021graphical};
however, in the next section we show that the full-data distribution is identifiable for the self-censoring model  characterized in Assumption \ref{assump:self} under a widely-used completeness condition.
The self-censoring model can be viewed as a generalization of  the shadow variable approach  \citep{miao2016on,d2010new} by admitting  missing outcomes as shadow variables and identifying multivariate missing outcomes in a unifying way.
Under the self-censoring model,  $Y_{-i}$ serves as a vector of  shadow variables for each $Y_i$, which is independent of $R_i$ conditional on $(X,Y_i,R_{-i}=1)$. 
Previous shadow variable approaches typically require the shadow variable to be   fully-observed;  
however, here  $Y_{-i}$ is prone to nonignorable missingness, which further complicates identification and inference in ways not addressed by existing shadow variable literature. We also propose a blockwise self-censoring model in the Supplementary Material, which allows arbitrary missingness mechanism within each block and a self-censoring mechanism across blocks.

Although we do not emphasize graphical concepts, graphical models can provide insight and intuition for identification and analysis for multivariate missing data.
The self-censoring model can also be encoded in a chain graph  $\mathcal{G} = (V, E)$,
where the set of vertices $V$  is comprised of   $(Y,R)$ and the set of edges $E$ contains    directed edges from each $Y_{i}$ to  $R_{i}$,   directed edges among  $Y$, and  undirected edges among $R$. The directed edges among $Y$ may be reversed without altering the self-censoring model. See \citet{lauritzen1996chain} for details of chain graphs.
Figure \ref{fig:M1} presents an example chain graph for a self-censoring model with three  outcomes.  We characterize the conditional independence by a chain graph instead of the DAG, because the   edges among $R$ are not allowed to be directed under the self-censoring assumption.
For strict self-censoring model, this chain graph reduces to the directed acyclic graph in Figure \ref{fig:M2}, where the edges among $R$ disappear.

\begin{minipage}{0.48\linewidth}
    \centering
    \begin{tikzpicture}[scale=0.6,
->,
shorten >=2pt,
>=stealth,
node distance=1cm,
pil/.style={
	->,
	thick,
	shorten =2pt,}]
\node (Y1) at (0,0) {$Y_1$};
\node (Y2) at (3,0) {$Y_2$};
\node (Y3) at (6,0) {$Y_3$};
\node (R1) at (0,2.5) {$R_1$};
\node (R2) at (3,2.5) {$R_2$};
\node (R3) at (6,2.5) {$R_3$};
\foreach \from/\to in {Y1/R1,Y2/R2,Y3/R3,Y1/Y2,Y2/Y3}
\draw [](\from) -> (\to);        
\draw [->] (Y1) to [out=-40,in=-140] (Y3);  
\foreach \from/\to in {R2/R3,R1/R2}
\draw [-]  (\from) -- (\to);        
\draw [-] (R1) to [out=40,in=140] (R3);  
\end{tikzpicture}
\captionsetup{justification=centering}

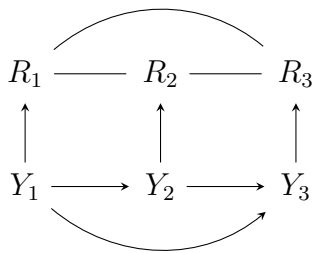
\captionof{figure}{Chain graph for  a self-censoring model.} \label{fig:M1}
\end{minipage}
\hfill
\begin{minipage}{0.48\linewidth}
    \centering
\begin{tikzpicture}[scale=0.6,
->,
shorten >=2pt,
>=stealth,
node distance=1cm,
pil/.style={
	->,
	ultra thick,
	shorten =2pt,}]

\node (Y1) at (0,0) {$Y_1$};
\node (Y2) at (3,0) {$Y_2$};
\node (Y3) at (6,0) { $Y_3$};
\node (R1) at (0,2.5) {$R_1$};
\node (R2) at (3,2.5) {$R_2$};
\node (R3) at (6,2.5) {$R_3$};
\foreach \from/\to in {Y1/R1,Y2/R2,Y3/R3,Y1/Y2,Y2/Y3}
\draw [](\from) -- (\to);        
\draw [->] (Y1) to [out=-40,in=-140] (Y3);  
\draw [-,color=white] (R1) to [out=40,in=140] (R3);  
\end{tikzpicture}
\captionsetup{justification=centering}

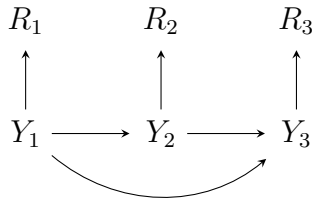
\captionof{figure}{Directed acyclic graph   for a strict self-censoring model.} \label{fig:M2}
\end{minipage}

\subsection{Identification}\label{subsec:22}
The joint distribution $\f(y, r \mid x)$ is identified if and only if the missingness mechanism $ \pr(r\mid x,y)$ is identified by  $\f(y, r \mid x)  = \f(y, r=1 \mid x) \pr(r\mid x,y)/\pr(r=1\mid x,y)$. To identify $\pr(r\mid x,y)$, we first split it into several parts by the odds ratio factorization \citep{chen2010compatibility, malinsky2021semiparametric} which holds under any missingness mechanism,
\begin{equation} 
	\pr(r\mid x,y)=\frac{\prod_{i=1}^{p} \pr\left(r_{i} \mid x,y, r_{-i}=1\right)\prod_{i=2}^{p} \eta_{i} \left(r_{i}, r_{<i}, x, y\right)}{\sum_{r'\in\mathcal{R}}\prod_{i=1}^{p} \pr\left(r^{\prime}_{i} \mid x,y, r_{-i}=1\right)\prod_{i=2}^{p} \eta_{i} \left(r'_{i}, r'_{<i}, x, y\right)},\label{eq:cond}
\end{equation}
where  $\eta_i$ is  the sequential odds ratio function between the $i$th and the preceding  missingness indicators,
\[\eta_{i} \left(r_{i}, r_{<i}, x, y\right)  = \frac{\pr(r_{i} \mid x,y, r_{<i}, r_{>i}=1)\pr(r_{i}=1 \mid x,y, r_{-i}=1)}{\pr(r_{i}=1 \mid x,y, r_{<i}, r_{>i}=1)\pr(r_{i} \mid x,y, r_{-i}=1)},\]
quantifying the  association between the missingness indicators; 
for  the strict self-censoring model,  $\eta_{i} \left(r_{i}, r_{<i}, x, y\right)=1$. Although we express the missingness mechanism by the sequential odds ratio factorization, the term $\prod_{i=2}^{p} \eta_{i} \left(r_{i}, r_{<i}, x, y\right)$ can be represented in terms of pairwise odds ratios and higher-order as in \citet{malinsky2021semiparametric}, thus not relying on particular ordering of the outcomes. 
In order to  identify  the joint distribution $\f(y, r\mid x)$, equality \eqref{eq:cond} shows that we only need to identify the itemwise propensity score  $\pr\left(r_{i} \mid x,y, r_{-i}=1\right)$ and the sequential odds ratio $\eta_i$ for all $i$. 
These quantities are not identified without model assumptions on the missingness mechanism. The  itemwise propensity score $\pr\left(r_{i} \mid x,y, r_{-i}=1\right)$ reduces to $\pr(r_{i}\mid x,y_{i}, r_{-i}=1)$ under Assumption \ref{assump:self}.
For notational  convenience, we let   $\pi_i(x, y_{i})$ denote $\pr(r_{i}=1\mid x, y_{i}, r_{-i}=1)$ and  $O_{i}(x, y_{i})$ denote the itemwise  odds function
\begin{equation}
	O_{i}(x, y_{i}) =  \frac{1-\pi_i(x, y_{i})}{\pi_i(x, y_{i})}. \label{eq:odds}
\end{equation}
We write   $\eta_i$, $\pi_i$  and $O_{i}$ for short where it causes no confusion.
Under the self-censoring model we have the following results that are useful for identification. 
\begin{theorem}\label{thm:identity}
Under  Assumption \ref{assump:self}, the sequential odds ratio function $\eta_{i}(\cdot)$ does not depend on $y$ and
	\begin{eqnarray}
	\eta_{i} \left(r_{i}, r_{<i}, x\right)  &=& \frac{\f(x, r_{\leq i},r_{>i}=1)}{\int \f(x, y_{i}, r_{<i}, r_{\geq i}=1)\cdot O_{i}(x, y_{i})^{1-r_i} d\nu(y_{i})}, \label{eq:seq}
	\end{eqnarray}
	for $i=2,\dots,p$.
\end{theorem}

Equation \eqref{eq:odds} and \eqref{eq:seq}  reveal the essential role of $O_{i}$ in identification of the joint distribution:
once $O_{i}$ is uniquely determined,   we can identify the  itemwise propensity score  $\pi_i$ and thus the sequential odds ratio $\eta_{i}$, which together suffice for identification of $\f(y,r\mid x)$. 
Besides, \eqref{eq:seq}   states that the sequential odds ratio $\eta_{i}$   is a function of $x, r_{\leq i}$ but   does not depend on $y$ under the self-censoring model. 
Next, we connect each $O_i$ to the observed-data distribution by
\begin{equation}
	E\{O_{i}(x,Y_{i}) \mid x, y_{-i}, r=1\} =  \frac{ \f(x, y_{-i}, r_{i}=0, r_{-i}=1)}{\f(x, y_{-i},r=1)}.\label{eq:intequ}
\end{equation}
Identification of $O_{i}$ is achieved as long as   the solution to equation \eqref{eq:intequ} is unique, which is a Fredholm integral equation of the first kind with $\f(x,y_{-i}, r_{i}=0, r_{-i}=1)$, $\f(x,y_{-i},r=1)$ and $\f(y_{i}\mid x, y_{-i}, r=1)$ obtained  from the observed-data distribution and $O_{i}(x ,y_{i})$ to be solved for. 
To identify $O_{i}$, we make the following completeness assumption.
\begin{assumption} \label{assump:cmp}
	 Completeness: $\f(y_{i}\mid x, y_{-i}, r=1)$ is complete in $y_{-i}$ for $i=1,\dots,p$ and any fixed $x$.
\end{assumption}
Completeness  is a  fundamental concept in statistics; see \cite{lehman1950completeness}. 
A conditional distribution  $\f(u\mid v)$ is \textit{complete} in $v$ if  $E\{g(U)\mid V\} = 0$ almost surely implies $g(u) = 0$ almost surely for any square-integrable function $g$.
Suppose $U,V$ are discrete variables with levels $u_k\in \{u_{1},\dots, u_{K}\}$ and $v_l\in \{v_{1},\dots, v_{L}\}$,
the completeness condition is equivalent to   full rank of matrix $\{\f(u\mid v)\}_{K\times L}$ with   elements $\f(u_k\mid v_l)$.
Completeness  is routinely assumed in nonparametric identification problems and is viewed as a regularity condition for identification, for example, in missing data \citep{d2010new, yang2019causal}, causal inference \citep{miao2018proxy, miao2021identifying, jiang2021identification}, measurement error \citep{an2012well}, and instrumental variable identification \citep{newey2003instrumental}; see \cite{miao2021identifying} for a  review. 
Assumption \ref{assump:cmp} implicitly requires  that   the support of $Y_{-i}$  has at least the same number of  levels   as that of $Y_{i}$.
The essential idea behind Assumption \ref{assump:cmp} is that the variability of $Y_{-i}$ can capture any infinitesimal variability of $Y_{i}$ in the complete cases.
We have the following identification result of full-data distribution.
\begin{theorem}\label{thm:idn}
Under Assumptions \ref{assump:self} and   \ref{assump:cmp}, $O_{i}(x, y_{i})$ is identified for all $i$ and therefore the joint distribution $\f(y, r\mid x)$ is identified.
\end{theorem}
Theorem \ref{thm:idn} establishes   identification of the full-data distribution $\f(y,r\mid x)$ for the self-censoring model under the completeness assumption, 
without imposing additional parametric model assumptions. 
As shown by \cite{mohan2021graphical} and \cite{nabi2020full},   without restricting the observed-data distribution,  identification of full-data distribution is impossible in the presence of    self-censoring. 
However, it is possible to recover the full-data distribution by jointly harnessing additional features of the data.
Assumption \ref{assump:cmp} characterizes such a feature   that ensures identification  under the self-censoring model. 
Our use of   completeness    is analogous to  its use in   previous identification problems.
Assumption \ref{assump:cmp} ensures uniqueness of the solution to  \eqref{eq:intequ}, i.e., identification of $O_i$.
After that, we can then identify  $\pi_i,\eta_i$ as in \eqref{eq:odds}--\eqref{eq:seq} and finally identify $\pr(r\mid  x,y)$ as in \eqref{eq:cond}.
Our identification strategy   views $Y_{-i}$ as a vector of shadow variables for $Y_i$, 
but in contrast to previous shadow variable approaches \citep{d2010new,wang2014instrumental,miao2019identification} that require a fully-observed shadow variable,
here we generalize the shadow variable framework by  exploiting incomplete variables $Y_{-i}$ as shadow variables for  each missing outcome $Y_i$ and accounting for associations between 
missingness of multivariate outcomes by identifying the sequential odds ratio $\eta_i$.

\section{Estimation}\label{sec:3}

\subsection{Parameterization}
Suppose we wish to make inference about  a full-data functional    $\psi$,
which   is defined as   the unique solution of a given   estimating equation $E\{m(X, Y;\psi)\} = 0$.
For instance, the outcomes mean $\psi=E(Y)$ corresponds to $m(x, y;\psi) = y -\psi$. 
Having established identification of the full-data distribution, we can in principle first estimate $\f(x, y, r=1),\f(x,y_{-i},r_i=0,r_{-i}=1)$ and then plug them into equation \eqref{eq:intequ}
to solve for the itemwise odds function $O_i$,  and finally obtain the joint distribution $\f(y,r\mid x)$ and any functional of interest.
Although the first step can be achieved by standard nonparametric estimation, the convergence rate is slow particularly when multivariate outcomes are 
concerned, and  the situation becomes worse if additional fully-observed covariates $X$ must be taken into account.
Besides,  it is challenging to solve equation \eqref{eq:intequ}  due to the ill-posedness of the Fredholm equations of the first kind \citep{carrasco2007, crucinio2021particle}, which  further complicates statistical inference.
Therefore, we consider a parameterization of the joint distribution and develop semiparametric estimators that only require the working models to be partially correct. The full-data distribution can be expressed by the odds ratio factorization,
\begin{equation}
	\f(y, r \mid x) = \frac{\OR(y, r \mid x) \f(y\mid r=1, x) \pr(r\mid x, y_0)}{\sum_{r'\in\mathcal{R}} E\left\{\OR(Y, r' \mid x)\mid r=1, x\right\}\pr(r'\mid x, y_0)},  \label{eq:joint}
\end{equation}
which is widely  used in missing data problems, see \cite{osius2004association,chen2007semiparametric,kim2011semiparametric,tchetgen2018Discrete, malinsky2021semiparametric} for examples.
Here $y=y_0$ is a reference value in the support of $Y$ chosen by the analyst and
\[\OR(y, r\mid x) =\frac{\pr(r \mid x, y) \pr(r=1 \mid x, y_0)}{\pr(r \mid x, y_0) \pr(r=1 \mid x, y)},\]
which is the  odds ratio function between the vectors $Y$ and $R$ conditional on   $X$, capturing the dependence of  the missingness indicators $R$ on the outcomes $Y$ owing to a shift of $Y$ from the reference value $y_0$.
Equation \eqref{eq:joint} reveals  a congenial specification  of the joint distribution with three  variationally independent components: the baseline outcome distribution $\f(y\mid x, r=1)$;
the odds ratio function $\OR(y, r \mid x)$, which can be factorized by a product of itemwise odds ratio $\Gamma_i(x, y_{i})=O_{i}(x, y_{i})/O_{i}(x, y_{0i})$,
\begin{equation}
		\OR(y,r\mid x) = \prod_{i=1}^p \Gamma_{i}(x, y_{i})^{1-r_{i}},\label{eq:orrelation}
\end{equation}
where $y_{0i}$ denotes the $i$th entry of $y_0$;
the baseline propensity score $\pr(r\mid x, y_0)$ for each missingness pattern $r$.
We consider  statistical inference of $\psi$ under partially correct specification of these three models. 
In the following, we let  $\Pi_{r}(x, y) = \pr(r\mid x, y)$ denote  the propensity score for missingness pattern $r$. 
We use $\hat{E}$ for the empirical mean.

\subsection{Inverse probability weighted estimator}
We first consider an inverse probability weighting approach that entails a parametric working model for the propensity score $\Pi_{r}(x, y)$ for each missingness pattern $r$.
We model   $\Pi_{r}(x, y)$  by specifying    parametric working models for the odds ratio function $\OR(y, r\mid x;\gamma)$ and     the baseline propensity score  $\Pi_{r}(x, y_0;\alpha)$,
which is equivalent to modeling each $\Gamma_i(x, y_{i};\gamma_{i})$, $\pi_i(x, y_{0i};\alpha_{1i})$, $\eta_{i} \left(r_{i}, r_{<i}, x;\alpha_{2i} \right)$ and evaluating 
$\Pi_{r}(x, y)$ according to \eqref{eq:cond}.
Let $\gamma = (\gamma_1,\dots,\gamma_p)$, $\alpha _{1}= (\alpha_{11},\dots,\alpha_{1p})$, $\alpha_2 = (\alpha_{22},\dots,\alpha_{2p})$, and $\alpha = (\alpha_1,\alpha_2)$ denote the   parameters for the working models. 
The itemwise propensity score is determined by the itemwise odds ratio function $\Gamma_i(x, y_{i};\gamma_{i})$ and itemwise baseline propensity score $\pi_i(x, y_{0i};\alpha_{1i})$ as follows,
\[\pi_i(x, y_i;\alpha_{1i}, \gamma_i) = \frac{\pi_i(x, y_{0i};\alpha_{1i})}{\pi_i(x, y_{0i};\alpha_{1i}) +\{1-\pi_i(x, y_{0i};\alpha_{1i}) \}\Gamma_{i}(x,y_{i};\gamma_i)}.\]

The estimation of $(\alpha, \gamma)$ is motivated by the following equations  that characterize   the propensity scores:
for $i=1,\ldots, p$ and $r\in \mathcal{R}$, 
$$
\begin{aligned}
	& E\left\{\frac{\mathbbm1(R=1)}{\pi_i(X,Y_i)} - \mathbbm1(R_{-i}=1)\mid X, Y_{-i}\right\} =0, \\
	& E\left\{\mathbbm 1(R=1)\cdot\frac{\Pi_{r}(X,Y)}{\Pi_{1}(X,Y)} - \mathbbm 1(R=r) \mid X\right\}\\
	=&E\left\{\mathbbm1(R=1)\prod_{j:r_{j}=0}\frac{1-\pi_j(X,Y_j)}{\pi_j(X,Y_j)}\prod_{i=2}^p\eta_{i} \left(r_{i}, r_{<i}, X \right) - \mathbbm1(R=r)\mid X\right\}=0,
\end{aligned}
$$
where $\mathbbm1(\cdot)$ denotes the indicator function.
The first equation is a consequence of self-censoring and the second equation echoes  the definition of $\Pi_r$ by equality \eqref{eq:cond}. 
We estimate $(\alpha, \gamma)$ by solving 
	\begin{align}
		& \hat{E}\left[\left\{\frac{\mathbbm1(R=1)}{\pi_i(\hat\alpha_{1i}, \hat\gamma_{i})}-\mathbbm1(R_{-i}=1)\right\}\cdot g_i(X, Y_{-i})\right] =0, \label{eq:ipwi}\\
		& \hat{E}\left[\left\{\mathbbm1(R=1)\prod_{j:r_{j}=0}\frac{1-\pi_j(\hat\alpha_{1j}, \hat\gamma_{j})}{\pi_j(\hat\alpha_{1j}, \hat\gamma_{j})}\prod_{i=2}^p\eta_{i} \left(r_{i}, r_{<i}, X;\hat\alpha_{2i} \right)-\mathbbm1(R=r) \right\}\cdot h_{i}(X) \right]=0, \label{eq:ipwalpha2}
	\end{align}
for all $i$, where $g_i$ and $h_i$ are user-specified vector functions of the same dimension as $(\alpha_{1i}^{\t}, \gamma_{i}^{\t})^{\t}$ and $\alpha_{2i}$, respectively. 
These two estimating equations only involve the observed data. We let $\logit(x) = \log \{ x/(1 - x) \}$ denote the logit transformation and $\expit(x)=\exp(x)/\{1+\exp(x)\}$ its inverse.
Specifically, if $\pi_i(x,y_{0i};\alpha_{1i}) = \expit\{\alpha_{1i}^{\t}x\}$, $\Gamma_i(x,y_i;\gamma_i)=\exp(\gamma_i y_i)$ and $\eta_i(r_i,r_{<i},x;\alpha_{2i}) = \exp\{\alpha_{2i}^{\t}x\}$ for $r_i=0$ and $r_{<i}\neq 1$, 
one natural choice for $h_i$ is   $h_i(x)= \partial \log \{\eta_i(r_{i},r_{<i},x;\alpha_{2i})\}/\partial\alpha_{2i}=x$. 
In equation \eqref{eq:ipwi},  $Y_{-i}$ is  observed for  $R_{-i}=1$  and thus  used as a proxy for  $Y_{i}$ to capture the variation of $Y_{i}$ that is not available from the pattern $R_{-i}=1$. One may choose $g_i(x,y_{-i}) = \partial \logit \pi_i(x,\overline{y}_{-i};\alpha_{1i}, \gamma_i)/\partial(\alpha_{1i}, \gamma_i)=(x^{\t},\overline{y}_{-i})^{\t}$ where $\overline{y}_{-i}$ is the average of outcomes in $y_{-i}$.
For equation \eqref{eq:ipwalpha2}, we propose a dynamic programming algorithm for solving these equations efficiently when the number of parameters is large in the Supplementary Material.

Let  $(\hat\alpha_{\rm ipw}, \hat\gamma_{\rm ipw})$ denote the nuisance estimators obtained from  \eqref{eq:ipwi}--\eqref{eq:ipwalpha2}. 
To estimate $\psi$, we solve the following estimating equation,
\begin{equation}\label{eq:estipw}
    \hat{E}\left\{\frac{\mathbbm1(R=1)}{\Pi_1(\hat\alpha_{\rm ipw},\hat\gamma_{\rm ipw})}m(X, Y;\hat\psi_{\rm ipw})\right\} = 0,
\end{equation}
where the full-data estimating equation is evaluated in the complete cases and the inverse propensity score weighting removes the selection bias
due to missing data.
We will show later that if $\OR(y, r\mid x;\gamma)$ and  $\Pi_r( x, y_0; \alpha)$ are correctly specified, then \eqref{eq:ipwi}--\eqref{eq:estipw} are unbiased estimating equations for $(\alpha,\gamma,\psi)$.

\subsection{Outcome regression/imputation-based estimator}
Alternatively, we can estimate $\psi$ by   imputing the missing values.
A commonly-used approach for imputation is   the exponential tilting or Tukey's representation \citep{kim2011semiparametric,franks2020flexible}:
for each missingness pattern $R=r'$,
\begin{equation}\label{eq:reg}
\f(y\mid x, r') = \frac{\OR(y, r'\mid x)\f(y\mid x, r=1) }{E\{ \OR(Y, r'\mid x)\mid x, r=1\}}.
\end{equation}
The imputation  entails  a baseline outcome model $\f(y\mid x, r=1; \beta)$ for the complete cases and an odds ratio model $\OR(y, r \mid x; \gamma)$. 
We solve the following   equations for $i=1,\ldots, p$ to obtain the nuisance estimators $(\beta, \gamma)$,
\begin{align}
    &\hat{E}\left\{\mathbbm1(R=1) \cdot \frac{\partial }{\partial \beta}  \log \f(Y\mid X, R=1; \hat\beta) \right\} = 0,  \label{eq:regb}\\
    &\hat{E}\left\{\mathbbm{1}(R_{i}=0, R_{-i}=1)\cdot \left[g_i(X, Y_{-i}) - E \{g_i(X, Y_{-i})\mid X, R_{i}=0, R_{-i}=1; \hat\beta, \hat\gamma_{i}  \}\right]\right\} = 0. \label{eq:regi}
\end{align}
The conditional expectation  $E(\cdot \mid R_{i} = 0, R_{-i}=1, X; \beta, \gamma_{i})$ is evaluated under \eqref{eq:reg} and $g_i$  is a user-specified function  of the same dimension  as $\gamma_{i}$. 
Estimation of $\beta$ only involves complete cases.
Estimation of the odds ratio parameter $\gamma$ in   \eqref{eq:regi} is motivated by the fact that $E \left[\mathbbm1(R_{i}=0)\cdot \left\{l(X,Y) - E\left\{l(X,Y)\mid R_{i}=0, R_{-i}=1, X\right\} \right\}\mid R_{-i}=1\right] = 0 $ for any function $l(x,y)$. 
Because $Y_{i}$  is missing for $R_{i}=0$, we replace $l(x,y)$ with  $g(x, y_{-i})$ where   $Y_{-i}$ is observed for  $R_{-i}=1$ and used as shadow variables for $Y_i$.

Let $(\hat\beta_{\rm reg}, \hat\gamma_{\rm reg})$ denote the nuisance estimators obtained from   \eqref{eq:regb}--\eqref{eq:regi}. 
An outcome regression/imputation-based estimator for $\psi$ is given by solving
\begin{equation}\label{eq:estreg}
\begin{aligned}
	\hat{E}&\left[\mathbbm1(R=1)\cdot m(X, Y;\hat \psi_{\rm reg}) \right.\\
&+ \left. \sum_{r\neq 1}\mathbbm1(R=r)\cdot E\left\{m(X, Y; \hat\psi_{\rm reg})\mid X, Y_{(r)},R=r; \hat\beta_{\rm reg}, \hat\gamma_{\rm reg})\right\}  \right] = 0,
\end{aligned}
\end{equation}
where $E\left\{\cdot \mid X, Y_{(r)},R=r; \beta, \gamma \right\}$ is evaluated according to  \eqref{eq:reg} to impute the missing values for pattern $R=r$. 
We will show later that if $\OR(y, r\mid x;\gamma)$ and     $\f(y\mid x, r=1;\beta)$ are correctly specified, then \eqref{eq:regb}--\eqref{eq:estreg} are unbiased estimating equations for $(\beta,\gamma,\psi)$.

\subsection{Doubly robust estimator}
However,  consistency of the inverse probability weighted and regression/imputation-based estimators  is no longer guaranteed if  any of the required working models is incorrect.
Therefore, we   construct an estimator    that combines both  approaches and achieves double robustness against misspecification of the working models.   
We estimate the nuisance parameters   by solving   \eqref{eq:ipwalpha2},   \eqref{eq:regb}, and \eqref{eq:scdri} together and let $(\hat\alpha_{\rm dr},\hat\beta_{\rm dr}, \hat\gamma_{\rm dr})$ denote the nuisance estimators.
\begin{equation}\label{eq:scdri}
\begin{aligned}
	    \hat{E}&\left(\left\{\frac{\mathbbm1(R=1)}{\pi_i(\hat\alpha_{i}, \hat\gamma_{i})}-\mathbbm1(R_{-i}=1)\right\}\right. \\
    	 &\cdot \left. \left[g_i(X, Y_{-i}) - E\left\{g_i(X, Y_{-i})\mid X, R_{i}=0, R_{-i}=1; \hat\beta, \hat\gamma_{i} \right\}\right] \right) =0.
\end{aligned}
\end{equation}

Given   $(\hat\alpha_{\rm dr},\hat\beta_{\rm dr}, \hat\gamma_{\rm dr})$,  the  doubly robust estimator of $\psi$ is given by    the solution to \eqref{eq:estdr}.
\begin{equation}\label{eq:estdr}
    \begin{aligned}
    \hat{E} &\left[\frac{\mathbbm1(R=1)}{\Pi_{1}(\hat\alpha_{\rm dr},\hat\gamma_{\rm dr})}m(X, Y;\hat\psi_{\rm dr})+\sum_{r\neq 1}\mathbbm1(R=r) E\left\{m(X, Y; \hat\psi_{\rm dr})\mid X,Y_{(r)},R=r; \hat\beta_{\rm dr}, \hat\gamma_{\rm dr}\right\} \right.\\
    &- \left. \frac{\mathbbm1(R=1)}{\Pi_{1}(\hat\alpha_{\rm dr},\hat\gamma_{\rm dr})}\sum_{r\neq 1}\Pi_{r}(\hat\alpha_{\rm dr},\hat\gamma_{\rm dr})E\left\{m(X, Y; \hat\psi_{\rm dr})\mid X, Y_{(r)},R=r; \hat\beta_{\rm dr}, \hat\gamma_{\rm dr}\right\} \right] = 0.
    \end{aligned}
\end{equation}
Equations  \eqref{eq:ipwalpha2} and   \eqref{eq:regb} have been used in the inverse probability weighted and outcome regression/imputation-based approaches for estimation of $\alpha$ and $\beta$, respectively,
while in contrast we use a different equation \eqref{eq:scdri} for estimation of $\gamma$ and  \eqref{eq:estdr} for estimation of $\psi$.
Equations \eqref{eq:scdri} and \eqref{eq:estdr}   have an augmented inverse probability weighting form where an augmentation term involving  the outcome regression is included to correct    the bias of  the inverse probability weighted estimation  when the baseline propensity score model is incorrect. 
The resultant  estimators $\hat\gamma_{\rm dr}$   and $\hat{\psi}_{\rm dr}$ are in fact doubly robust. 
Let $\mathcal{M}_{r,i}$ and $\mathcal{M}_{y,i}$ denote the following semiparametric models,
\begin{itemize}
    \item[] $\mathcal{M}_{r,i}$: $\pi_i(x, y_{0i}; \alpha_{1i})$ and $\Gamma_i(x, y_{i};\gamma_{i})$ are correctly specified; 
    \item[] $\mathcal{M}_{y,i}$: $\f(y \mid x, r = 1; \beta)$ and $\Gamma_i(x, y_{i};\gamma_{i})$ are correctly specified.
\end{itemize}
The following theorem summarizes   properties  of  the odds ratio estimators.
\begin{theorem}\label{thm:drgamma}
    Under Assumptions \ref{assump:self} and \ref{assump:cmp}, and the regularity conditions in the Supplementary Material, 
    then
    \vspace{-5pt}
\begin{enumerate}
\setlength{\itemsep}{-25pt}
\item[(i)]   $\hat{\gamma}_{i,\rm ipw}$ is consistent and asymptotically normal in model $\mathcal{M}_{r,i}$;\\
\item[(ii)]$ \hat{\gamma}_{i,\rm reg}$ is consistent and asymptotically normal in model $\mathcal{M}_{y,i}$;\\
\item[(iii)]$\hat{\gamma}_{i, \rm dr}$ is consistent and asymptotically normal in the union model $\mathcal{M}_{r,i}\cup \mathcal{M}_{y,i}$.
\end{enumerate}
\end{theorem}
Let $\mathcal{M}_{r} = (\cap_{i=1}^p\mathcal{M}_{r,i})\cap (\cap_{i=2}^p\{\eta_{i}(r_{i}, r_{<i}, x;\alpha_{2i}) \text{ is correct}\})$ and $\mathcal{M}_{y} = \cap_{i=1}^p\mathcal{M}_{y,i}$, 
that is,
    \begin{itemize}
        \item[]  $\mathcal{M}_{r}$: $\Pi_r( x, y_0; \alpha)$ and $\OR(y, r \mid x;\gamma)$ are correctly specified;
        \item[]  $\mathcal{M}_{y}$: $\f(y \mid x, r = 1;\beta)$ and $\OR(y, r \mid x;\gamma)$ are correctly specified.
    \end{itemize}
The following theorem summarizes   properties  of  the  estimators of $\psi$.
\begin{theorem}\label{thm:drpsi}
    Under Assumptions \ref{assump:self} and \ref{assump:cmp}, and the regularity conditions in the Supplementary Material, 
    then
    \vspace{-5pt}
\begin{enumerate}
\setlength{\itemsep}{-25pt}
\item[(i)]   $\hat{\psi}_{\rm ipw}$ is consistent and asymptotically normal in model $\mathcal{M}_{r}$;\\
\item[(ii)] $\hat{\psi}_{\rm reg}$ is consistent and asymptotically normal in model $\mathcal{M}_{y}$;\\
\item[(iii)] $\hat{\psi}_{\rm dr}$ is consistent and asymptotically normal in the union model $\mathcal{M}_{r}\cup \mathcal{M}_{y}$.
\end{enumerate}
\end{theorem}

Doubly robust methods have been promoted for missing data analysis, causal inference and many other  coarsened data problems, see \cite{tsiatis2007semiparametric} and \cite{seaman2018introduction} for a review. 
Theorems \ref{thm:drgamma}  and \ref{thm:drpsi} show the double robustness   of $\hat\gamma_{i,\rm dr}$ and $\hat{\psi}_{\rm dr}$.
The   consistency   of $\hat\gamma_{i,\rm dr}$   requires correct specification of the itemwise odds ratio $\Gamma_i(x, y_{i};\gamma_{i})$   and  of  at least one of the baseline models  $\{\pi_i(x, y_{0i}; \alpha_{1i}), \f(y \mid x, r = 1;\beta)\}$.
The consistency of $\hat{\psi}_{\rm dr}$ requires stronger conditions---correct specification of $\OR(y, r \mid x;\gamma)$,  i.e., all   the itemwise odds ratio  models and correct specification of at least one of $\{\Pi_r( x, y_0; \alpha), \f(y \mid x, r = 1;\beta)\}$.
The doubly robust estimators   offer  one more chance to correct  bias due to partial model misspecification;
although, they  will generally also be biased if both baseline models are incorrect \citep{kang2007demystifying} or the odds ratio function is incorrect.
Following the general theory for estimation equations \citep{newey1994asymptotic,van2000asymptotic}, variance estimation of the estimators can be obtained and   confidence intervals can be constructed based on the normal approximation, which is described in the Supplementary Material. One can also  implement the   bootstrap method.

\section{Simulation Study}\label{sec:4}
We evaluate the   performance of the proposed estimation methods  via simulations. 
One fully-observed covariate $X$   is generated from ${\rm Unif}(-1,1)$ and three continuous outcome variables $(Y_1,Y_2,Y_3)$ and missingness indicators $R$ are generated  from the self-censoring model, where    we consider four scenarios  for $\f(y,r\mid x)$ with  different choices for the baseline propensity score  and baseline outcome distribution. The detailed data generating mechanism is described in the Supplementary Material. 
The parameter of interest is the outcome  mean  of $Y_1$, i.e., $\psi=E(Y_1)$.

Let   $X_1=(1,X)^\t$. For estimation, we specify a  multivariate normal distribution $N(\beta^\t x_{1}, \Sigma)$   for the baseline outcome distribution $\f(y\mid x,r=1)$
and a bilinear  model  \citep{chen2004nonparametric}  for the odds ratio function $\OR(y, r\mid x) = \exp\{\sum_{i=1}^3\gamma_i (1-r_i) y_i\}$.  
The working model for the baseline propensity score $\Pi_r( x, y_0; \alpha)$ is specified with  a logistic model for each itemwise baseline  propensity score $\pi_i(x, y_{0i})=\expit(\alpha_{1i}^\t x_{1})$ and a  bilinear  model   for   the sequential odds ratios $\eta_2=\exp\{\alpha_{21}(1-r_1r_2) x\}, \eta_3=\exp\{\alpha_{22}(1-r_3)(1-r_1r_2) x\}$. 
We apply  the  three proposed estimators and compare them to a benchmark doubly robust estimator obtained by assuming missing at random.
We also  compute   variances of these estimators and  evaluate the coverage rate of the Wald-type  confidence interval.

\begin{figure}[ht]
	\centering
	\subfloat[TT]{
		\includegraphics[width=.24\textwidth,height=.28\textwidth]{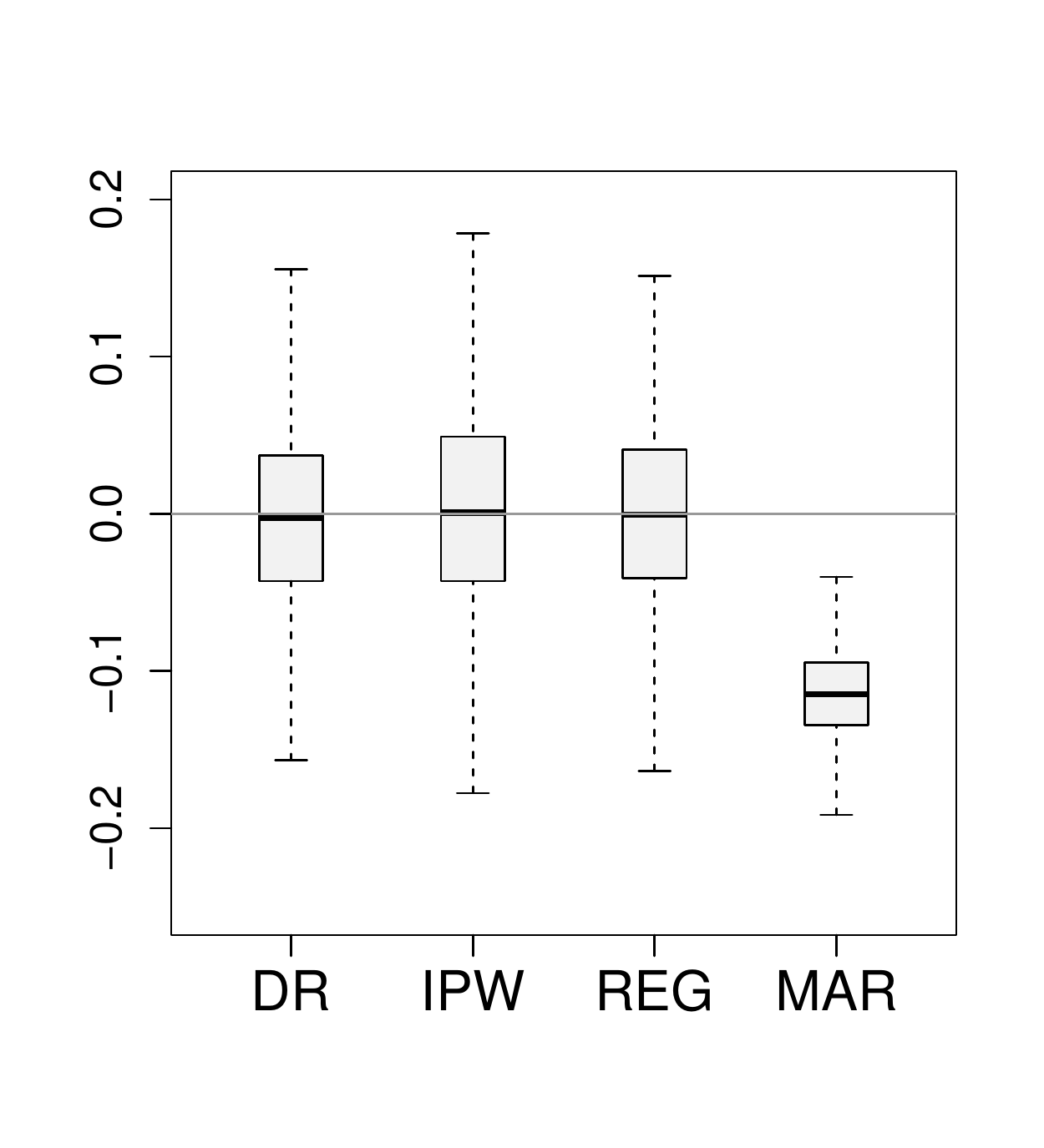}  \label{psiFT}}
	\subfloat[TF]{
		\includegraphics[width=.24\textwidth,height=.28\textwidth]{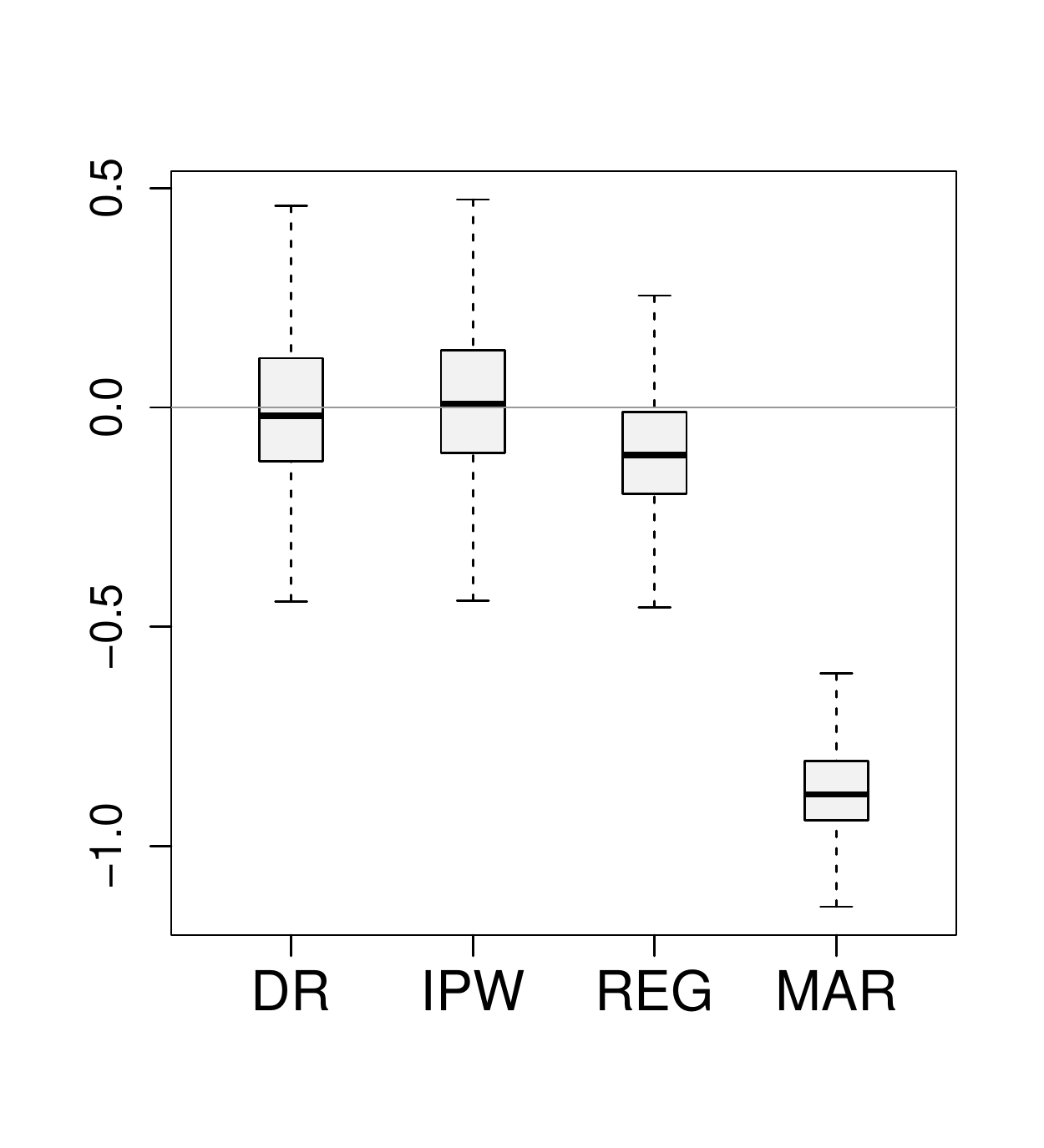}  \label{psiTF}}
	\subfloat[FT]{
		\includegraphics[width=.24\textwidth,height=.28\textwidth]{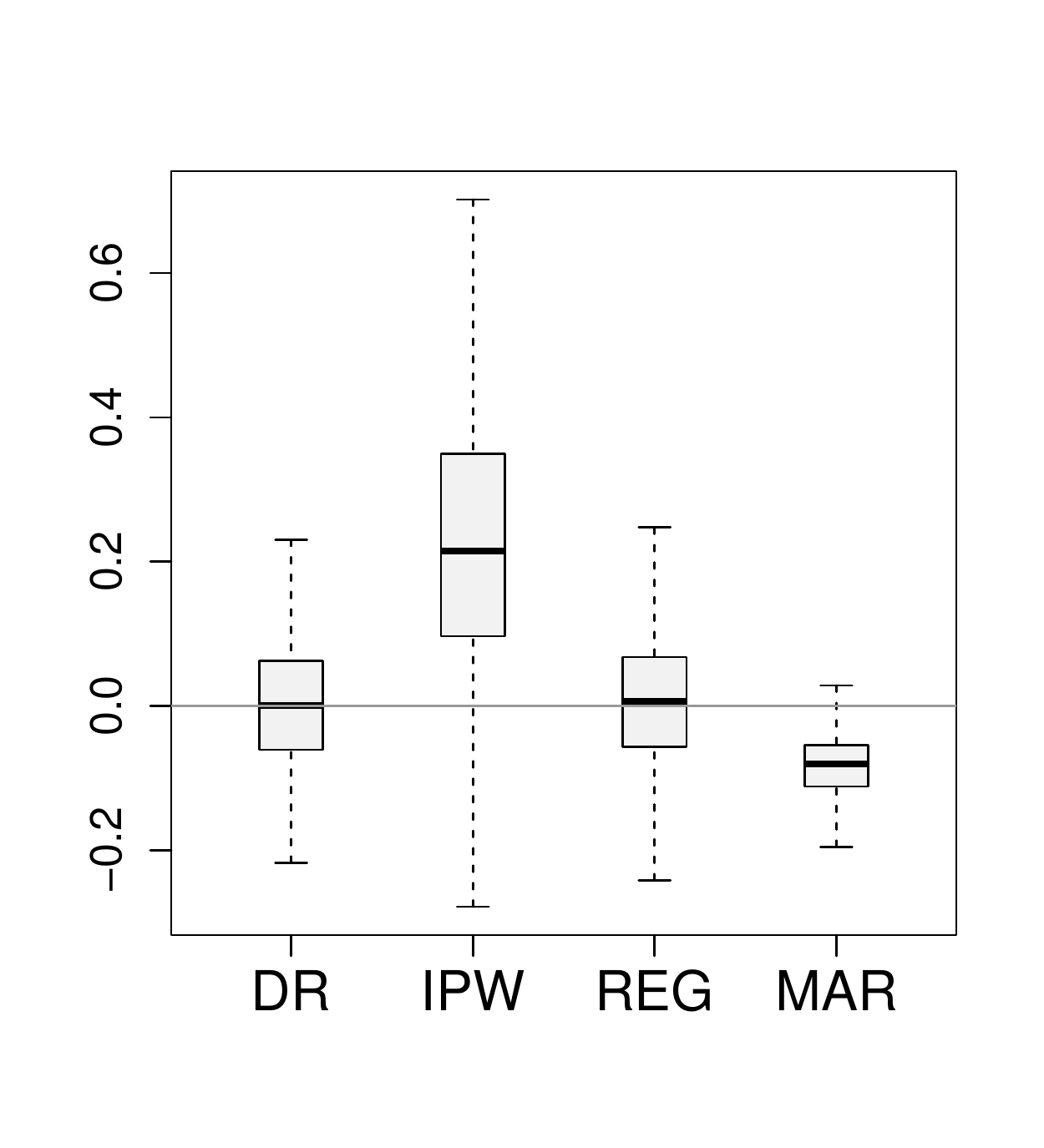}  \label{psiTT}}
	\subfloat[FF]{
		\includegraphics[width=.24\textwidth,height=.28\textwidth]{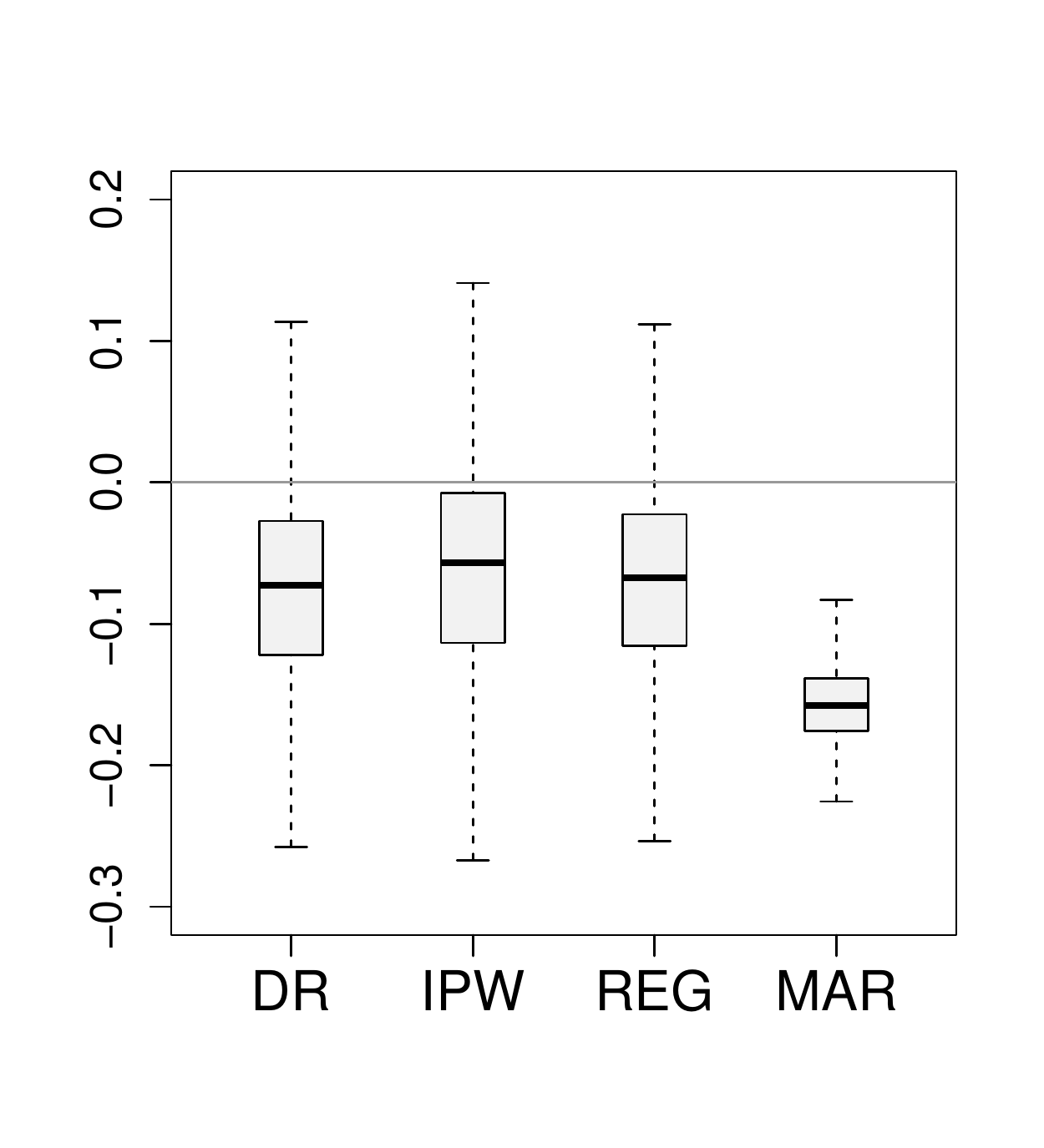}  \label{psiFF}}
	\caption{Bias of  the outcome mean estimators.
	} \label{fig:simu1mu}
	\centering
	\subfloat[TT]{
		\includegraphics[width=.24\textwidth,height=.28\textwidth]{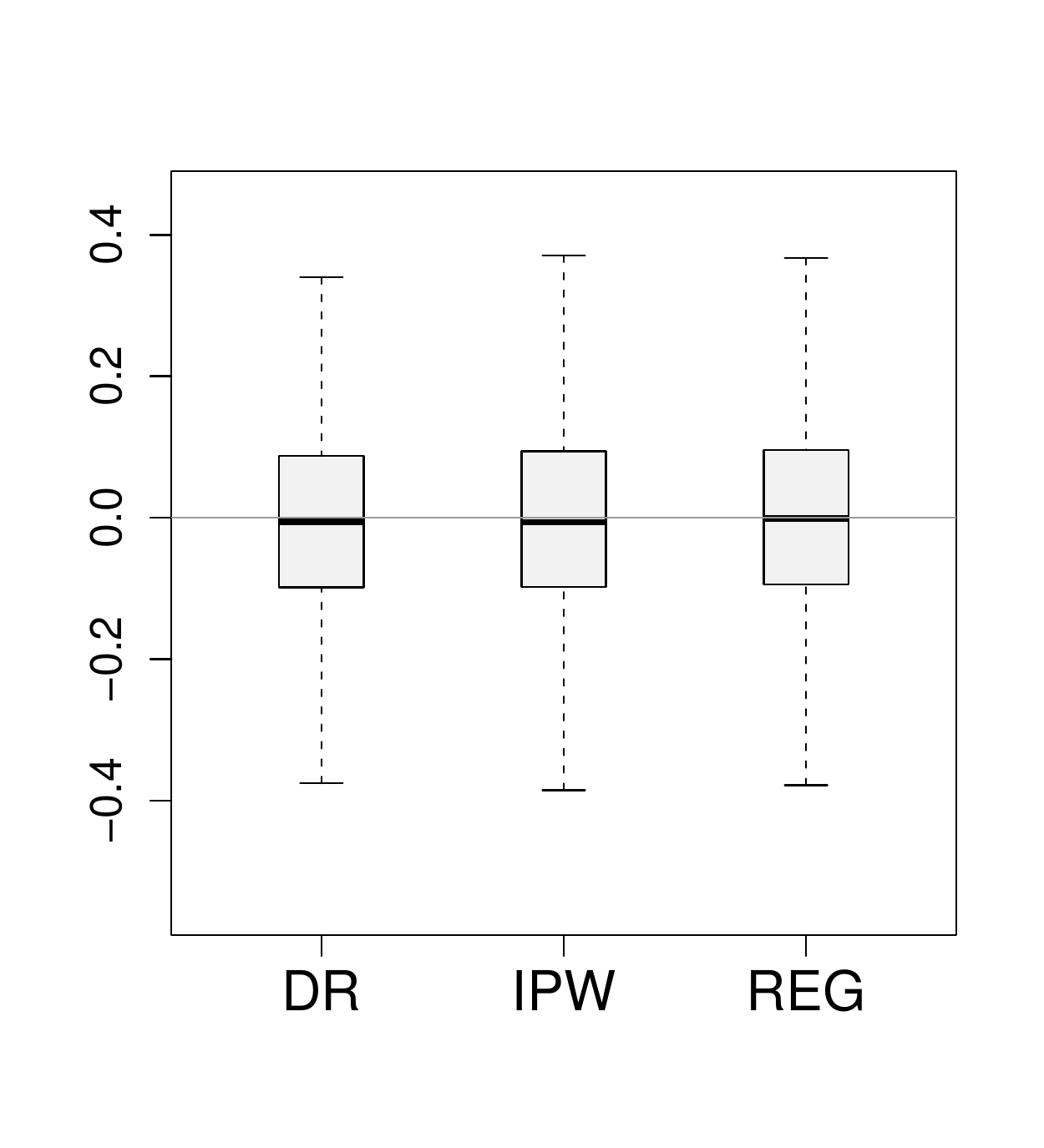} \label{gammaFT}}
	\subfloat[TF]{
		\includegraphics[width=.24\textwidth,height=.28\textwidth]{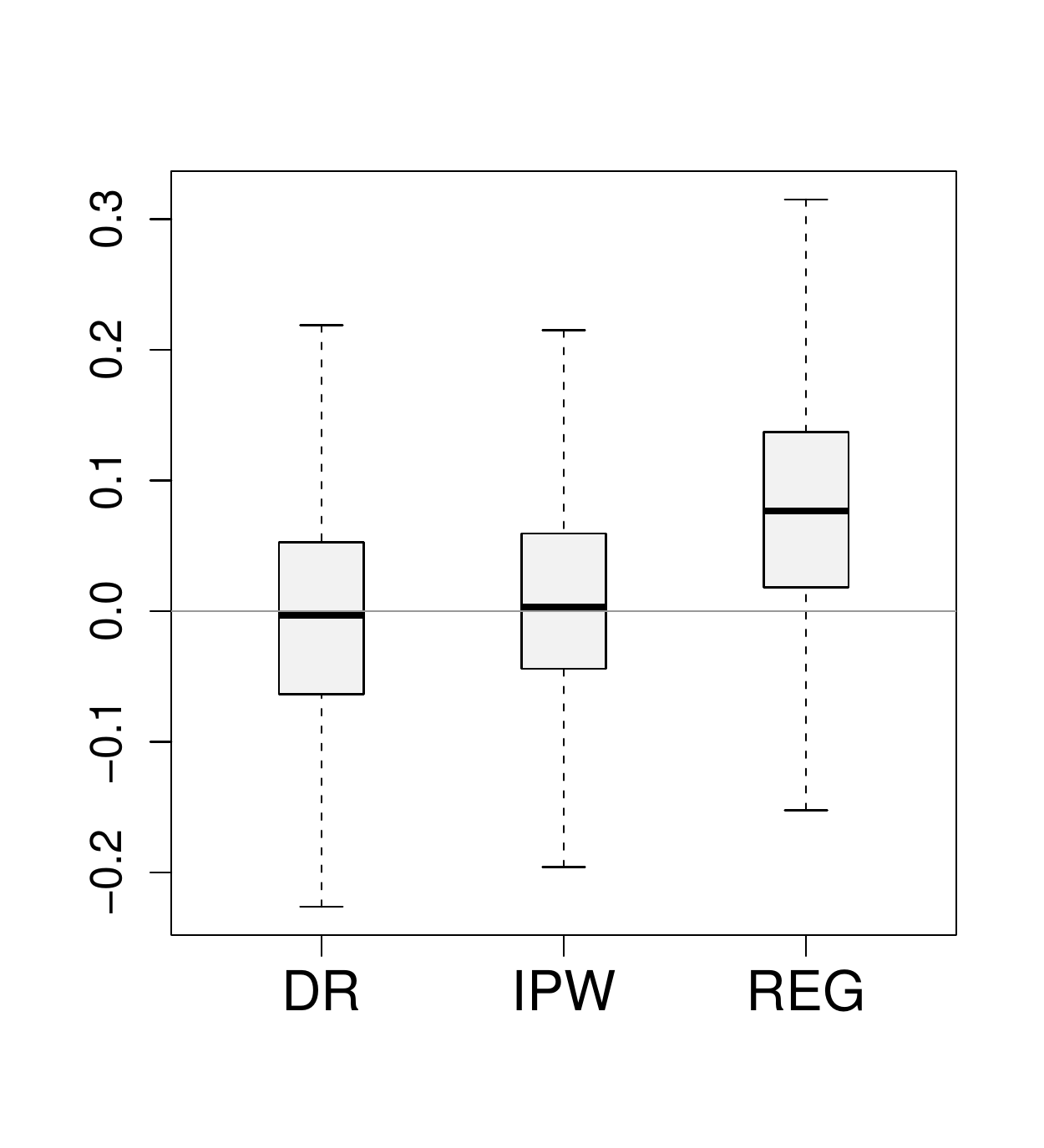}\label{gammaTF}}
	\subfloat[FT]{
		\includegraphics[width=.24\textwidth,height=.28\textwidth]{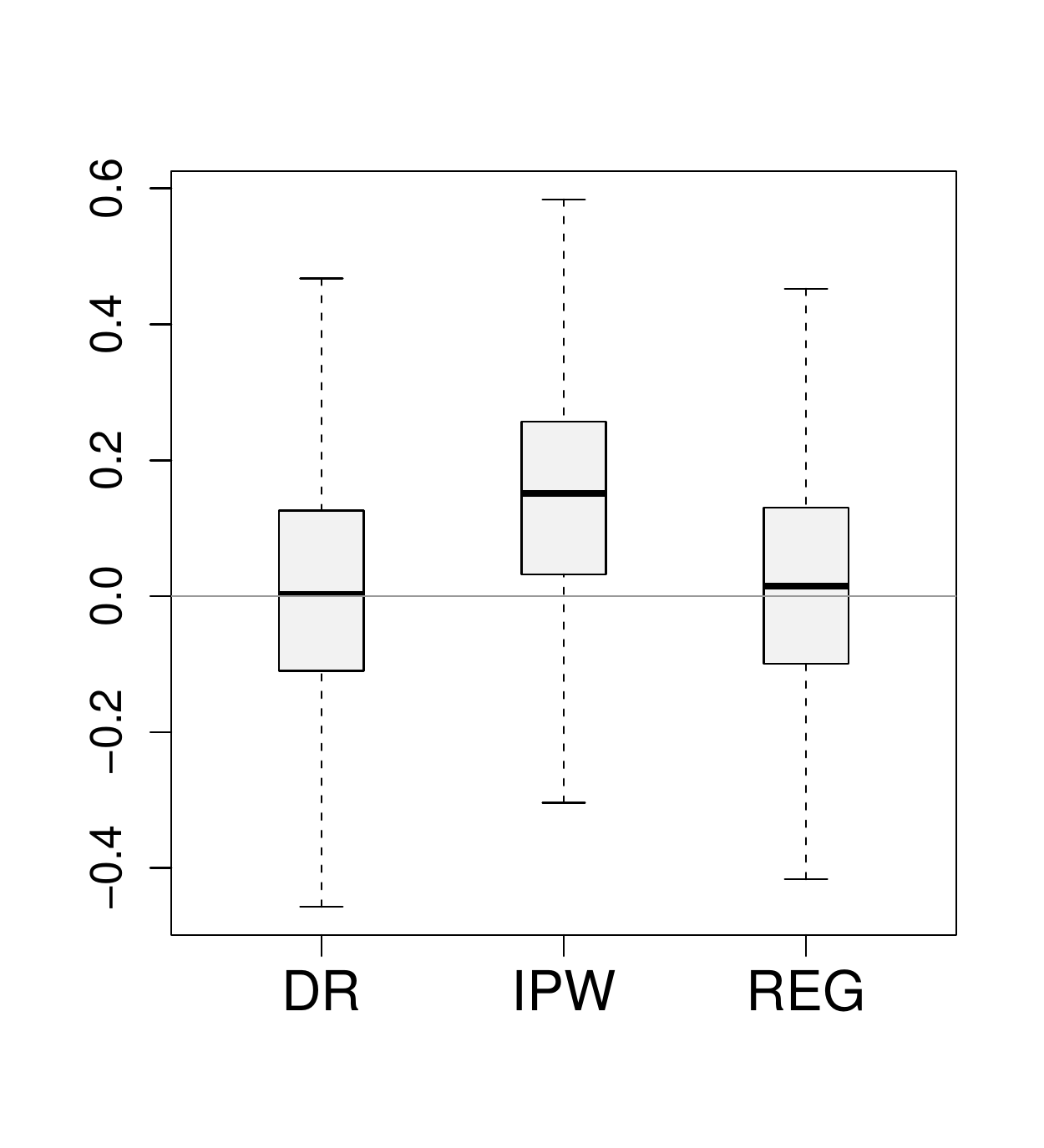}\label{gammaTT}}
	\subfloat[FF]{
		\includegraphics[width=.24\textwidth,height=.28\textwidth]{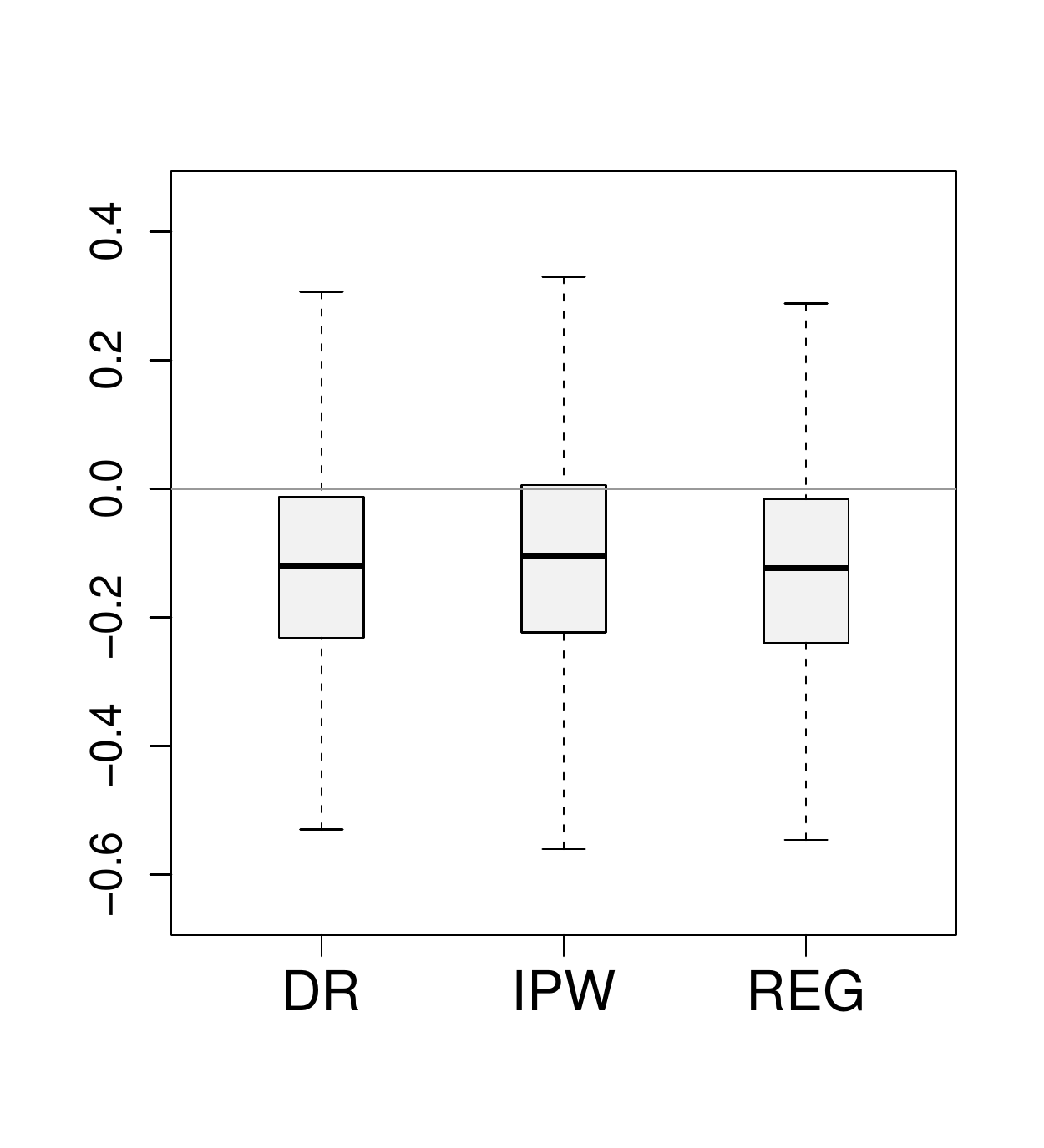}\label{gammaFF}}
	\caption{Bias of  the odds ratio estimators.} \label{fig:simu1or}
\end{figure}

\begin{table}[htp]
	\center
	\caption{\textit{Coverage probability of the 95\% confidence interval}} \label{tbl:cvrn}
	\begin{tabular}{lcccccccccccccccccclllll}
		&\multicolumn{7}{c}{Outcome mean ($\psi$)} & & & \multicolumn{5}{c}{Odds ratio parameter ($\gamma_{1}$)} \\
		 &  DR & & IPW & & REG && MAR  && & DR && IPW & & REG  \\
		TT &0.964&& 0.968&& 0.940&& 0.044&&& 0.958&& 0.957 &&0.958\\
		TF &0.933&& 0.963&& 0.878&& 0.095&&& 0.935&& 0.949 &&0.861\\
		FT &0.969&& 0.771&& 0.958&& 0.394&&& 0.958&& 0.882 &&0.958\\
		FF &0.858&& 0.912&& 0.817&& 0.227&&& 0.869&& 0.888 &&0.874\\
	\end{tabular}
\end{table}

We replicate $1000$ simulations for each  scenario with sample size $n=3000$ and summarize the results for estimation bias of  the outcome mean $\psi$ and odds ratio parameter $\gamma_1$  in Figure \ref{fig:simu1mu} and \ref{fig:simu1or}, respectively.  
Table \ref{tbl:cvrn} shows   coverage rates of the $95\%$ confidence interval. 
In scenario TT, all working models are correct, then the three proposed estimators show little bias  and the coverage rate of the $95\%$ confidence interval is  close to 0.95.
In scenario TF, the baseline propensity score model $\Pi_r( x, y_0; \alpha)$ and the odds ratio model  $\OR(y, r\mid x;\gamma)$ are correct but the baseline outcome model  $\f(y\mid x,r=1;\beta)$ is incorrect, then the inverse probability weighted and doubly robust estimators show little bias with fine coverage rates, while the regression/imputation-based estimator has  large bias with low coverage rate. 
In scenario FT,  the baseline  propensity score model $\Pi_r( x, y_0; \alpha)$   is incorrect but  the baseline outcome model $\f(y\mid x,r=1;\beta)$ and the odds ratio model $\OR(y, r\mid x;\gamma)$ are correct, then the inverse probability weighted estimator has non-negligible bias with a coverage rate well below 0.95 while the regression/imputation-based and doubly robust estimators have little bias with coverage rates close to 0.95. 
However, when both baseline working models are incorrect, all three  proposed estimators are biased. 
As expected, the conventional   doubly robust estimator obtained under missing at random has  large bias in all scenarios  with  a poor coverage rate. 
In summary, we recommend the doubly robust estimator for multivariate missing data subject to self-censoring.

\section{Real data analysis}\label{sec:5}

We analyze a dataset extracted from an observational study of HIV-positive mothers in Botswana. It is of interest to evaluate the association between maternal highly active antiretroviral therapy in pregnancy and preterm delivery adjusted for CD$4^+$ lymphocyte cell count.
Details of the study  are described in \cite{chen2012highly}. 
We focus on   $n=2341$ HIV-positive women with maternal hypertension.
Our   analysis includes three binary variables, namely an indicator of continuing highly active antiretroviral therapy during pregnancy (70.3\% missing), 
an indicator of preterm delivery (15.5\% missing), and an indicator of whether CD$4^+$ cell count is less than 200 $\mu$L (62.2\% missing). The dataset only has a small proportion of complete cases (8.33\%) and the missingness is nonmonotone.
Our primary interest is the risk difference of highly active antiretroviral therapy on preterm delivery  after adjustment for two CD$4^+$ cell count levels. 
Let ${\rm RD}_1,{\rm RD}_2$ denote   the risk differences in the  low and normal  CD$4^+$ count groups, respectively.

We apply the   self-censoring model and implement the   proposed methods for estimation.
We compare our methods to the augmented inverse probability weighted estimator under no self-censoring model (NSC) from \citet{malinsky2021semiparametric} and multivariate imputation by chained equations (MICE) from \citet{van2011mice} where each   incomplete outcome is imputed with a univariate logistic regression model assuming missing at random and  10 imputed datasets are pooled to obtain the estimates.
To apply  the proposed estimation methods,  we  use  multinomial distributions for  the baseline propensity score $\Pi_r(y_0)$ and the baseline outcome models $\f(y\mid r=1)$ because the outcomes are binary,
and we model the odds ratio with $\OR(y, r) = \exp\{\sum_{i=1}^3\gamma_i (1-r_i) y_i\}$, where each $\gamma_i$ characterizes the extent of nonignorable missingness of each outcome.
The 95\% confidence intervals are obtained with bootstrap.

\begin{table}[ht]
	\center
	\caption{\textit{Point estimate and 95\% confidence interval of odds ratio parameters}} \label{tbl:odds}
	\begin{tabular}{lcccccccccccccc}
		Method &&  \multicolumn{2}{c}{$\gamma_1$} &&  \multicolumn{2}{c}{$\gamma_2$}  && \multicolumn{2}{c}{$\gamma_3$}  \\
		IPW && -0.10&(-1.36,0.97)&& -3.32 &(-6.30,-2.09)&&-1.03&(-2.28,0.83)\\
		REG && 0.01&(-0.23,0,46)&& -5.32&(-8.97,-4.25) &&-1.19&(-5.13,0.09)\\
		DR && 0.06&(-0.43,0.58)&& -5.10&(-8.03,-3.12) &&-1.06&(-3.24,0.14)\\
	\end{tabular}
\end{table}
\begin{table}[ht]
	\center
	\caption{\textit{Point estimate and 95\% confidence interval  of risk differences}} \label{tbl:rd2}
	\begin{tabular}{lcccccccc}
		Method &\multicolumn{2}{c}{${\rm RD}_1$} &&&  \multicolumn{2}{c}{${\rm RD}_2$} \\
		IPW  &0.293 & (-0.042, 0.649) &&& 0.156  & (-0.013, 0.305)\\
		REG &0.266  & (-0.040, 0.586) &&& 0.151  &(-0.009, 0.301)\\
		DR    &0.267  &(-0.042, 0.597) &&& 0.152   &(-0.011, 0.302)\\
		NSC   &0.289  &(-0.046, 0.699) &&& 0.184  &(0.046, 0.315)\\
		MICE &0.097 &(-0.028, 0.427)&&& 0.060	&(-0.148, 0.174) 
	\end{tabular}
\end{table}

Table \ref{tbl:odds} shows the estimates of the odds ratio parameters $\gamma$. 
The point estimation of $\gamma_1$ is close to zero, suggesting that  the missingness of  preterm delivery is likely    to be ignorable.
However, the estimates of $\gamma_2$ and $\gamma_3$ are significantly negative.
This is evidence for  nonignorable missingness and suggests that    mothers with low CD$4^+$ count or receiving highly active antiretroviral therapy are  more likely   to  respond. 
Table \ref{tbl:rd2} reports  the estimates of the risk differences ${\rm RD}_1$ and ${\rm RD}_2$. 
Our analysis results based on the self-censoring model show  that the use of highly active antiretroviral therapy contributes to this risk of preterm delivery for mothers with HIV-positive and maternal hypertension.
All three estimates obtained under the self-censoring model agree that highly active antiretroviral therapy may lead to a higher risk of  preterm delivery. We also compare our estimates with the results based on the no self-censoring assumption. The conclusion is consistent in the way that both self-censoring and no self-censoring models detect significant risk differences for the two groups.
In contrast, the multiple imputation method  substantially underestimates the two risk differences, because it does not account for nonignorable missingness of low CD$4^+$ count and highly active antiretroviral therapy.

\section{Discussion}\label{sec:6}

It is   of interest to develop an efficient estimator that attains the semiparametric efficiency bound in the union model $\mathcal M_r\cup \mathcal M_y$.
Yet, due to the complexity of the multivariate self-censoring model, a closed expression of such an efficient influence function and efficient estimator is currently not available.
Moreover,  they involve complex features of the observed data distribution which are difficult to model correctly, and thus the potential prize  of   implementing the efficient estimator may not always be worth the chase \citep{stephens2014locally}.

The proposed identification and estimation can be extended to a  blockwise self-censoring model: 
the missing outcomes  can be partitioned into different blocks so that   the missingness  is arbitrary within each block but the missingness of each block is not affected by other blocks.
Figure \ref{fig:M4} presents an example chain graph for such a model  where the edges from outcome block $(Y_{i1},Y_{i2})$ to missingness indicator block $(R_{j1},R_{j2})$ are not present for $i\neq j$.
Identification  of the  blockwise self-censoring model is  established under a blockwise completeness condition  in the Supplementary Material.
It is also  of interest to study the identification and inference of a mixed-censoring model that consists of both self-censoring    and no self-censoring.
Figure \ref{fig:M3} presents an example  chain graph for such a model where (i) the study variables can be grouped into two disjoint sets $(W,Y)$ with missingness indicator sets $(M,R)$; 
(ii) the missingness of $W$ is no self-censoring and of $Y$ is self-censoring, i.e., $M_i\ind W_i\mid (W_{-i},M_{-i}, Y, R)$ for $i=1,\dots,p$ and $R_{j}\ind (W,M,Y_{-j})\mid (Y_{j},R_{-j})$ for $j=1,\dots,q$; 
(iii)   $M$ and $R$ are independent conditional on $(W,Y)$.
Then we can  identify $\f(y, r)$ by applying the proposed identification  strategy  under the completeness assumption and identify $\f(w, m\mid y, r=1)$ by applying the identification results of the no self-censoring model  \citep{sadinle2017itemwise,malinsky2021semiparametric}. 
Finally, the full-data distribution $\f(w,y, m,r)$ is identified   because $\f(w,m\mid y,r=1)=\f(w,m\mid y, r)$. 
However,  the identification of the mixed-censoring and the blockwise self-censoring model becomes challenging if one does not have prior knowledge about the self-censoring and no self-censoring cliques or the independence assumptions encoded by the graphical structures. 
We plan to  study statistical inference and their connections to  the graphical models  in the future. 

\begin{minipage}[c]{\linewidth}
\centering
\begin{tikzpicture}[scale=0.65,
->,
shorten >=2pt,
>=stealth,
node distance=1cm,
pil/.style={
	->,
	thick,
	shorten =2pt,}]

\node (Y1) at (0,0) {$Y_1$};
\node (Y2) at (3,0) {$Y_2$};
\node (Y3) at (6,0) {$W_1$};
\node (Y4) at (9,0) {$W_2$};
\node (R1) at (0,2.5) {$R_1$};
\node (R2) at (3,2.5) {$R_2$};
\node (R3) at (6,2.5) {$M_1$};
\node (R4) at (9,2.5) {$M_2$};
\foreach \from/\to in {Y1/R1,Y2/R2,Y2/Y1,Y3/Y2,Y2/R4,Y3/R4,Y1/R3,Y3/R4,Y4/Y3,Y4/R3}
\draw (\from) -- (\to);        
\draw [->] (Y3) to [out=-140,in=-40] (Y1);  
\draw [->] (Y4) to [out=-140,in=-40] (Y2);  
\draw [-] (R1) to  (R2); 
\draw [-] (R3) to  (R4);  
\end{tikzpicture}

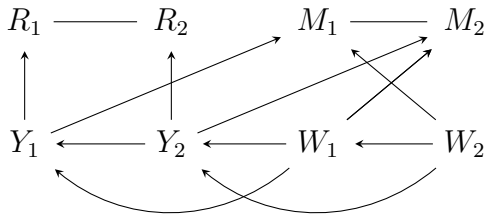
\captionof{figure}{Chain graph  for a  mixed-censoring model.} \label{fig:M3}
\end{minipage}

\begin{minipage}[c]{\linewidth}
\centering
\begin{tikzpicture}[scale=0.65,
->,
shorten >=2pt,
>=stealth,
node distance=1cm,
pil/.style={
	->,
	thick,
	shorten =2pt,}]

\node (Y1) at (0,0) {$Y_{11}$};
\node (Y2) at (3,0) {$Y_{12}$};
\node (Y3) at (7,0) {$Y_{21}$};
\node (Y4) at (10,0) {$Y_{22}$};
\node (Y5) at (14,0) {$Y_{31}$};
\node (Y6) at (17,0) {$Y_{32}$};
\node (R1) at (0,2.5) {$R_{11}$};
\node (R2) at (3,2.5) {$R_{12}$};
\node (R3) at (7,2.5) {$R_{21}$};
\node (R4) at (10,2.5) {$R_{22}$};
\node (R5) at (14,2.5) {$R_{31}$};
\node (R6) at (17,2.5) {$R_{32}$};
\foreach \from/\to in {Y1/R1,Y2/R2,Y2/Y1,Y3/Y2,Y4/R4,Y3/R4,Y1/R2,Y3/R4,Y4/Y3,Y4/R3,Y5/R5,Y5/R6,Y5/Y4,Y6/Y5}
\draw (\from) -- (\to);        
\draw [->] (Y3) to [out=-140,in=-40] (Y1);  
\draw [->] (Y4) to [out=-140,in=-40] (Y2);  
\draw [->] (Y6) to [out=-140,in=-40] (Y4);  
\draw [->] (Y5) to [out=-140,in=-40] (Y3);  
\draw [-] (R1) to  (R2); 
\draw [-] (R3) to  (R4);  
\draw [-] (R5) to  (R4);  
\draw [-] (R2) to  (R3); 
\draw [-] (R1) to [out=40,in=140] (R3);   
\draw [-] (R2) to [out=40,in=140] (R4);  
\draw [-] (R3) to [out=40,in=140] (R5);  
\end{tikzpicture}

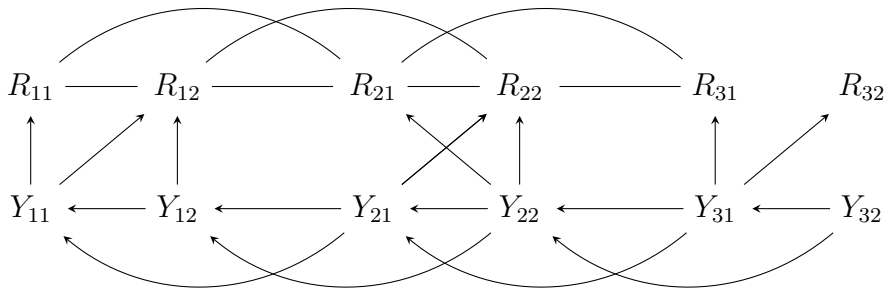
\captionof{figure}{Chain graph for a blockwise self-censoring  model.} \label{fig:M4}
\end{minipage}

\section*{Supplementary Material}
Supplementary material available online includes additional examples, a dynamic programming algorithm for solving equation \eqref{eq:ipwalpha2}, proof of theorems, simulation details, and identification for the blockwise self-censoring model.

\bibliographystyle{apalike}
\bibliography{CausalMissing}

\end{document}